\documentclass[a4paper,10pt]{article}
\usepackage[latin1]{inputenc}
\usepackage{amsmath}
\usepackage{graphicx}
\usepackage{epsfig}
\usepackage{float}
\usepackage{amsfonts}
\begin{document}

\begin{center}
{\Large\bf Quintessence Scalar Field: A Dynamical Systems Study}
\\[15mm]
Nandan Roy \footnote{E-mail: nandan@iiserkol.ac.in} and
Narayan Banerjee \footnote{E-mail: narayan@iiserkol.ac.in}

{\em Department of Physical Sciences,~~\\Indian Institute of Science Education and Research-Kolkata,~~\\Mohanpur Campus, West Bengal 741252, India.}\\[15mm]
\end{center}

\begin{abstract}
The present work deals with a dynamical systems study of quintessence potentials leading to the present accelerated expansion of the universe. The principal interest is to check for late time attractors which give an accelerated expansion for the universe. Two examples are worked out, namely the exponential and the power-law potentials.

\end{abstract}


\section{Introduction:}

That the universe expands with an acceleration is now an observational certainty\cite{r1a,r1b,r1c,r1d,r1e,r1f,r1g,r1h}. Although a new incarnation of the cosmological constant ${\Lambda}$ appears to be a good enough option as the driver of this acceleration, it has its own problems, particularly that of the compatibility of the theoretically predicted value and the observational requirement. There is a number of exhaustive reviews in the literature\cite{r3a,r3b,r3c}. The second most talked about option, a quintessence scalar field, provides an effective negative pressure $p_{\phi} = \frac{1}{2} \dot{\phi}^2 - V(\phi)$, which drives the acceleration. There is a huge variety of quintessence potentials which can serve this purpose \cite{r6,r7,r8,r9,r10,r11,r12,r13}. It deserves mention that none of the quintessence potential warrants an inevitable existence from theoretical consideration and none perhaps has a very strong observational advantage over the others. This actually explains the existence of so many possible potentials which are not yet completely ruled out. Some excellent reviews can be found in ref. \cite{r4a,r4b}. \\

The present work investigates about the possible stable solutions for some quintessence models. The field equations are written in terms of dimensionless quantities as an autonomous system. The relevant fixed points are located and the physically relevant solutions are looked at. The interest, of course, is to look for models that give an accelerated expansion for the present universe, but a relevant past is also considered in the discussion of numerical solutions that are obtained. \\

A dynamical systems study has aleardy found some applications in cosmology. For a systematic discussion on the early work in this direction, we refer to \cite{r17,r18}. A dynamical systems analysis for a scalar field in an FRW unverse had been discussed by Gunzig at al\cite{r19} and Carot and Collinge\cite{r20}. Both these investigations are regarding scalar fields giving rise to an inflationary scenario in the early universe. In the context of a late surge of the accelerated expansion of the universe, Urena-Lopez\cite{r21} employed this technique in the case of a scalar field leading to a phantom behaviour. Writing the system of equations as an autonomous system has its merits as one deals with first order equations. This method has been employed for an axionic quintessence by Kumar, Panda and Sen\cite{r22} and for the dynamics of a thawing dark energy by Sen, Sen and Sami\cite{r23}. Very recently a dynamical systems study of Phantom, Tachyonic and K-essence fields have been given by Fang et at \cite{r31}. They also discuss about a non-canonical quintessence field. \\

In a recent work, the stability of cosmological solutions for two quintessence potentials under the tracking condition has been studied\cite{r24}. Under the tracking condition, the quintessence field energy density ``tracks'' the dark matter density for the major part of the evolution but finally takes over to dominate the dynamics of the universe in the later stages of the evolution\cite{r25,r26,r27,r28}. \\

In the present work, the tracking condition is relaxed, and the stable solutions for two well known quintessence potentials are looked at (for a comprehensible list of quintessence potentials, we refer to \cite{r4b}). The autonomous system, written for a general potential $V=V(\phi)$, turns out to be a three dimensional system. For the exponential potential, the system practically reduces to a two-dimensional one. We look for numerical solutions for the system, and check their stability. For a power-law potential, the system is actually three dimensional. The numerical solutions are looked at, as usual, in keeping with the observational requirement of a universe undergoing an accelerated expansion in the later stages. In the latter case, as the system is 3-dimensional, the stability is analyzed with a different strategy, namely by a small perturbation around the fixed point.

In the second section we write down the system of equations as an autonomous system and discuss about the boundary conditions. The third section deals with the general method of the stability analysis. Section 4 deals with the actual examples of the two potentials mentioned and the fifth and final section includes a discussions on the results obtained.

\section{The dynamical system:}
We consider a spatially flat Robertson Walker universe, given by the metric
\begin{equation} \label{eq.metric}
ds^2 = dt^2 - a(t)(dr^2 + r^2 d\Omega ^2).
\end{equation}
The matter sector is considered to be a barotropic fluid with an equation of state $p _{B}=(\gamma-1) \rho_{B}$, where $\gamma$ is a constant, $\rho_{B}$ is the energy density of the perfect fluid and $p_{B}$ is the corresponding pressure. The subscript $B$ stands for the fact that the fluid is dominated by baryonic dark matter. In order to facilitate the late time accelerated expansion of the universe, a scalar field distribution is also considered. Einstein field equations are written as
\begin{equation} \label{H}
H^2= \frac{8 \pi G}{3}(\rho_{B} + \frac{1}{2} \dot{\phi}^2 + V(\phi)),
\end{equation}
and
\begin{equation} \label{eq.dH}
\dot{H}=-\frac{8 \pi G}{2}(\gamma \rho_{B} + \dot{\phi}^2).
\end{equation}
Here $H=\frac{\dot{a}}{a}$ is the Hubble parameter, $\phi$ is the scalar field and $V(\phi)$ is the scalar potential. A dot represents a differentiation with respect to time $t$.  The conservation equation for the fluid is
\begin{equation} \label{eq.rho}
\dot{\rho}_{B}= -3 \gamma H \rho_{B}.
\end{equation}
The wave equation is given by
\begin{equation} \label{eq.phi}
\ddot{\phi} + 3 H \dot{\phi} = - \frac{dV}{d\phi}.
\end{equation}

Not all these equations are independent as the wave equation can be derived from the other three in view of Bianchi identities. Equations  (\ref{eq.dH}), (\ref{eq.rho}), and (\ref{eq.phi}) are chosen to be the equations constuting the system of equations to be solved, subject to a constraint equation(\ref{H}). We define dimensionless variables,

\begin{equation} \label{eq.trans}
x^2=\frac{k^2 {\phi^{\prime}}^2}{6}, ~~ y^2 = \frac{k^2 V}{3 H^2} ,
\end{equation}
where a prime denotes a differentiation with respect to $N=ln(\frac{a}{a_0})$ and $k^2=8 \pi G$. The present value of the scale factor, $a_0$, is subsequently chosen as unity. The energy density and the effective pressure due to the scalar field $\phi$ can be written respectively as 
\begin{equation} \label{eq.rhop}
\rho_{\phi}= \frac{1}{2} {\dot{\phi}}^2 + V(\phi),~~ p_{\phi}= \frac{1}{2} {\dot{\phi}}^2 - V(\phi).
\end{equation}
These two can be formally connected by an equation of state 
$p_{\phi}=(\gamma_{\phi}-1) \rho_{\phi}.$
Thus the equation of state parameter $\gamma_{\phi}$ for the scalar field can be written as 
\begin{equation} \label{eq.gamma}
\gamma_{\phi}=\frac{\rho_{\phi} + p _{\phi}}{\rho_{\phi}} = \frac{{\dot{\phi}}^2 }{\frac{{\dot{\phi}}^2}{2} + V} = \frac{2 x^2}{x^2 + y^2}.
\end{equation}
The system of equations can be written in terms of these new variables as a 3-dimensional autonomous system (with N as the argument),
 \begin{equation} \label{eq.x}
 x^{\prime} = -3x + \lambda \sqrt{\frac{3}{2}} y^2 +\frac{3}{2} x [2 x^2 + \gamma (1-x^2-y^2)]~~~,
\end{equation}

\begin{equation} \label{eq.y}
y^{\prime} =- \lambda \sqrt{\frac{3}{2}} xy + \frac{3}{2} y [2 x^2 + \gamma (1-x^2-y^2)] ~~~,
\end{equation}
and 
\begin{equation} \label{eq.lambda}
\lambda^{\prime} = -\sqrt{6} \lambda^2 (\Gamma -1) x ~~~,
\end{equation} \\
where $ \lambda = - \frac{1}{k V} \frac{dV}{d \phi}$ and $\Gamma =  V\frac{d^2 V}{d \phi^2} / (\frac{dV}{d \phi})^2$. \\
The density parameter $\Omega_{\phi}$ for the scalar field is given by 

\begin{equation} \label{eq.omega}
\Omega_{\phi}=\frac{k^2 \rho_{\phi}}{3 H^2}= x^2 + y^2,
\end{equation}
which, in view of the constraint equation (\ref{H}), is restricted as 
\begin{equation}
 \Omega_{B} + \Omega_{\phi}=1 , 
 \end{equation}
 for a spatially flat universe, where $\Omega_{B}$ is baryonic energy density parameter. This implies $0\leq \Omega_{\phi} \leq 1$. 
\par
From equation (\ref{eq.dH}), the deceleration parameter ($q = -\frac{a \ddot{a}}{a^2} = - \frac{\dot{H} + H^2}{H^2}$ ) can be expressed in terms of $x$ and $y$ as

\begin{equation} \label{eq.q}
q= \frac{3 \gamma}{2} (1- x^2 -y^2) + 3x^2 -1.
\end{equation}

In order to find the numerical solutions for the system one has to fix the boundary values. We estimate the boundary values of $x, y$ etc using the present ( i.e., at $N = ln a = 0$) values of  $\Omega_{B}, q$ etc as suggested by observations. Present observational value of $\Omega_{B}= 0.27$, so
\begin{equation} \label{eq.boun}
 \Omega_{\phi} = x_{0}^2 + y_{0}^2=0.73 , 
 \end{equation}
  where $x_{0}$ and $y_{0}$ are the present values of $x$ and $y$ respectively. Using equation ({\ref{eq.boun}}) in equation ({\ref{eq.q}}), a lower bound is set for present value of deceleration parameter ($q_{0} > -0.595$) so as to make $x_{0}$ real. 
  \\
  If we set $q_{0}=-0.53$, consistent with observations (see ref.\cite{r29} and references therein) and the lower bound set for the present work, then equations (\ref{eq.q}) and (\ref{eq.boun}) would yield $x_{0}=0.147$ and $y_{0}=0.842$. It is easy to check that $\Gamma = 1$ corresponds to an exponential potential and other values of $\Gamma$ correspond to power law type potentials. The present value of $\lambda$ has been chosen to fit the observational results. In this work we have checked for numerical solutions that evolve from a fixed point(asymptotic as $N \longrightarrow -\infty$) of the system and attracted towards  another fixed point (asymptotic as $N \longrightarrow \infty$) of the system. Thus, the solutions are heteroclinic orbits in phase space which connects two different fixed points. From this behaviour of the numerical solutions, the beginning and the ultimate fate of the universe can be qualitatively investigated. It deserves mention here that in keeping with the requirement that the dark matter is expected to be cold, the subsequent discussion assumes that the corresponding pressure $p_{B}$ to be zero which implies $\gamma =1$.

\section{Stability analysis:}
Let us consider a system of differential equations given by\\
$x^\prime=f(x)$,\\
where $x^\prime = \frac{dx}{dt}, x \equiv (x_{1},x_{2},....x_{n}) \in \mathbb{R}^n $ and $f:\mathbb{R}^n \longrightarrow \mathbb{R}^n $.\\
Fixed points of the system are the points $x_{0} \in \mathbb{R}^n$ such that $f(x_{0}) = 0$. A non linear system can be linearised at fixed points as \\
$u^\prime = A u = Df(x_{0})u $,\\
where $u=x-x_{0}, A = Df(x_{0})$ and $Df(x)= \frac{\partial f_{i}}{\partial x_{j}},$ $i,j= 1,2,....n$.
 A is called the Jacobian matrix for the system. If the fixed point is hyperbolic then the stability of the fixed points can be determined from the eigen values of the Jacobian matrix at that fixed point by using Hartman-Grobman theorem ( see, any standard text on Dynamical Systems e.g.\cite{r30}). If all the  eigen values have negative real part then the fixed point is stable i,e. future time attractor. If all eigen values have positive real part then the fixed point is unstable i,e. past time attractor. If eigen values have both positive and negative real parts then the fixed point is saddle. For a 2-dimensional system, the stability of a fixed point can also be determined from determinant($\bigtriangleup$) and trace ($\tau$) of the Jacobian matrix(A) at that fixed point. If $\tau<0,\bigtriangleup>0$, then the fixed point is stable. For $\tau>0$, $\bigtriangleup>0$ and $\bigtriangleup<0$ indicates an ustable and a saddle type fixed points respectively.

\section{Examples with specific potentials:}

\textbf{(i) Exponential Potential:}
 When $\Gamma =1$ the potential has the form 
\begin{equation}\label{exp.pot}
 V(\phi) = A e^{\alpha \phi},
\end{equation}
 where A and $\alpha$ are constants. From equation (\ref{eq.lambda}), $\lambda={\lambda}_{0}$ is a constant. So the problem effectively becomes a 2-dimensional one. With a numerical solution of the equations (\ref{eq.x}) and (\ref{eq.y}), the plots for $x$ and $y$ are obtained. Plots for observationally relevant cosmological parameters have also been obtained. For these plots, we have taken $\lambda_{0} =1$. This also yields $\alpha=-k$. The analysis is possible for some other values of $\lambda$, but there is hardly any qualitative difference, so we do not include them.\\
 
\begin{figure}[H]
\centering
\includegraphics[scale=0.5]{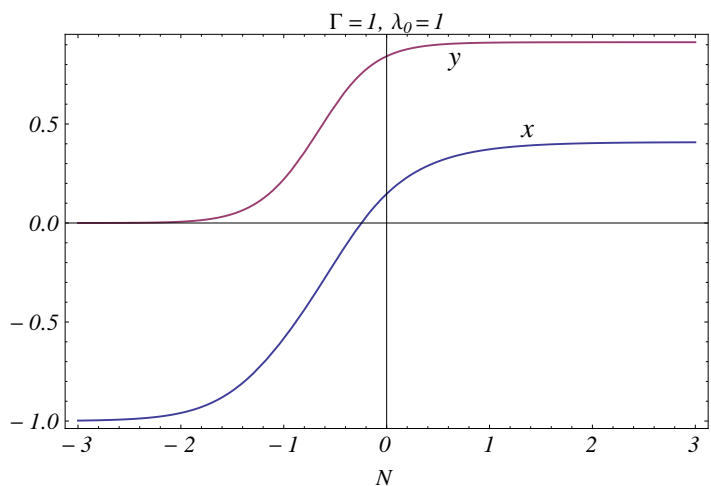}
\caption{The plot of $x$ and $y$ vs $N$ for exponential potential.}
\end{figure}
~\\
Figure 1 is the plot of $x,y$ against $N$. In figure 2, cosmological parameters $q, \Omega_{\phi}$ and $\Omega_{B}$ are shown  against N.
The fixed points($x_{0},y_{0}$) of the system when $\lambda_{0} = 1$ and $\gamma=1$ are (1.224,-1.224), (0.408,-0.912), (1,0),   (-1,0), (1.224,1.224), (0.408,0.912) and (0,0). From figure 1, we see that ($x,y$) is asymptotic to (-1,0) as $N \longrightarrow -\infty$ in the past and to (0.408,0.912) as $N \longrightarrow \infty$ in the future. Determinant($\bigtriangleup$) and trace($\tau$) of the Jacobian matrix at (-1,0) are $\bigtriangleup = 3 (3 +\sqrt{\frac{3}{2}})$ and $\tau = 6 + \sqrt{\frac{3}{2}}$. As both these are positive, the point (-1,0) is an unstable fixed point. Determinant($\bigtriangleup$) and trace($\tau$) of the Jacobian matrix at (0.408,0.912) are $\bigtriangleup = 5.123$ and $\tau =-4.55$, so this point is a stable fixed point. The fixed point (-1,0), which is past time attractor of our solution has the qualitative features like $a \longrightarrow 0, \phi \longrightarrow \infty, V(\phi) \longrightarrow 0, H \longrightarrow \infty $ as $N \longrightarrow -\infty$. The qualitative features of the fixed point (0.408,0.912), which is a future time attractor, are $a \longrightarrow \infty, \phi \longrightarrow \infty, V(\phi) \longrightarrow 0$ and $q \longrightarrow -0.5$ as $N \longrightarrow \infty$. \\

This analysis, along with reasonable boundary values (i.e. those indicated by observations), shows that the present accelerated expansion is indeed a stable solution for the universe with an exponential potential for the quintessence scalar field. The fixed point $(x,y)=(-1,0)$ is unstable, and thus, given a perturbation, the universe evolves from that state when the linear size ($a$) is zero, the value of the scalar field infinity and the potential was zero. It starts expanding very rapidly ($H \longrightarrow \infty$) but at a decelerated rate. After  this phase the universe has a transition to an accelerated expansion phase which is shown in figure 2. The accelerated expansion starts at about $Z \simeq 0.49$,  which is quite consistent with the observations. From figure 2 , it is also seen that in recent past $\Omega_{\phi} < \Omega_{B}$, but in remote past, $\Omega_{\phi}$ dominates over $\Omega_{B}$ though at that time the universe was undergoing a decelerated expansion.This is indeed intriguing.  We see that in the remote past $\gamma_{\phi} > 1$ ( figure 3) and the pressure and the density of the quintessence field are related as $p_{\phi} = (\gamma_{\phi}-1) \rho_{\phi}$. This indicates that the contribution to the pressure sector from the quintessence field had been positive and hence the quintessence field failed to drive an acceleration. But as the universe evolves, the value of $\gamma_{\phi}$ drops below unity and the effective pressure becomes negative so as to drive an accelerated expansion. \\

The future behaviour of the universe in this model can be derived qualitatively analysing the fixed point (0.408,0.912). This fixed point is a future time attractor as discussed before. So as $N \longrightarrow \infty$, the universe settles to a state where ${\Omega}_{B} \longrightarrow {0}$ and ${\Omega}_{\phi}$ attains aconstant value. The universe expands with a nearly constant acceleration.

\begin{figure}[H]
\centering
\includegraphics[scale=0.5]{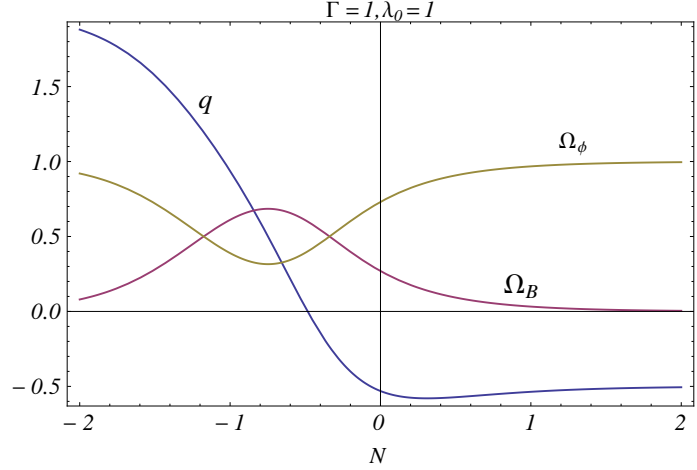}
\caption{The plot of $q, \Omega_{B}$ and $\Omega_{\phi}$ vs $N$ for exponential potential.}
\end{figure}

\begin{figure}[H]
\centering
\includegraphics[scale=0.5]{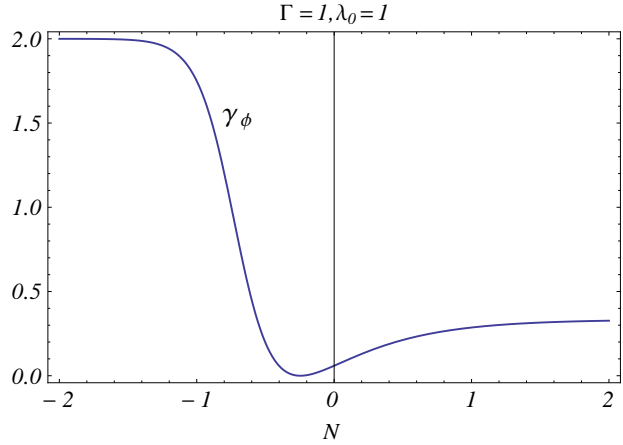}
\caption{The plot of $\gamma_{\phi}$ vs $N$ for exponential potential.}
\end{figure}

\textbf{(ii) Power law type potentials:}  
When $\Gamma =m$, a constant, the form of the potential can be found out from definition of $\Gamma$ as
\begin{equation}\label{pwrlaw.pot}
 V(\phi) = (A \phi + B)^{\frac{1}{1-m}},
\end{equation}
  where A,B and m are constants. In this case $\lambda$ is not necessarily a constant, and we plot $x,y,\lambda$ and also the cosmological parameters like $q, \Omega_{B}, \omega_{\phi}, \gamma_{\phi}$, all against $N$. As an example, we take $\Gamma = 4$. The fixed points ($x_{0},y_{0},\lambda_{0}$) of the system for power law type potentials are (-1,0,0), (0,0,0), (1,0,0), (0,-1,0) and (0,1,0). These fixed points are independent of $\Gamma$. From figure 4 we see that ($x,y,\lambda$) curves are asymptotic to (-1,0,0) as $N \longrightarrow -\infty$ and asymptotic to (0,1,0) as $N \longrightarrow \infty$. The eigen values of the Jacobian of the system at (-1,0,0) is (3,3,0) and those at (0,1,0) is (-3,-3,0). As these fixed points are non-hyperbolic, we can not use linear stability analysis. The fixed point (-1,0,0) has two positive eigenvalue and one zero eigenvalue so it is an unstable fixed point. However, the fixed point (0,1,0) has two negative eigen value and one zero eigenvalue, so 
 its being a stable fixed point cannot be ruled out before further investigation. \\
In order to check the stability of the latter fixed point,  we perturb the system in every direction near the fixed point and numerically solve the system of equations to find its asymptotic behaviour as $N \longrightarrow \infty$. The plots for this numerical perturbation are given in figure 7, 8 and 9. The 3D phase portrait looks obscure and it is difficult to figure out the behaviour. So we have plotted $x, y$ and $\lambda$ against $N$ in figure 7,8 and 9. From figure 7, we see that $x(N)$ approaches $x=0$ as $N \longrightarrow \infty$. Also from figure 8 and figure 9, one concludes that y(N) approaches $1$ and $\lambda(N)$ approaches $0$ respectively as $ N \longrightarrow \infty$. So the system approaches the fixed point (0,1,0) as $N \longrightarrow \infty$ against perturbation near (0,1,0). The fixed point is thus  a stable one. Thus, for a power law type potential also, a qualitative analysis can be looked at. \\

The universe evolves from the unstable fixed point (-1,0,0) representating the physical state ($a \longrightarrow 0, \phi \longrightarrow \infty, V(\phi) \longrightarrow$ constant and finte, $H \longrightarrow \infty $). As it is an unstable state, a small perturbation starts the evolution. After a long stint of decelerated expansion, the system enters into the phase of  accelerated expansion and is attracted towards the stable fixed point (0,1,0). This final stable state will correspond to the physical state ($a \longrightarrow \infty, \phi \longrightarrow$ constant, $ V(\phi) \longrightarrow$ constant) and it will expand with a nearly constant acceleration (figure 5).

\begin{figure}[H]
\centering
\includegraphics[scale=0.5]{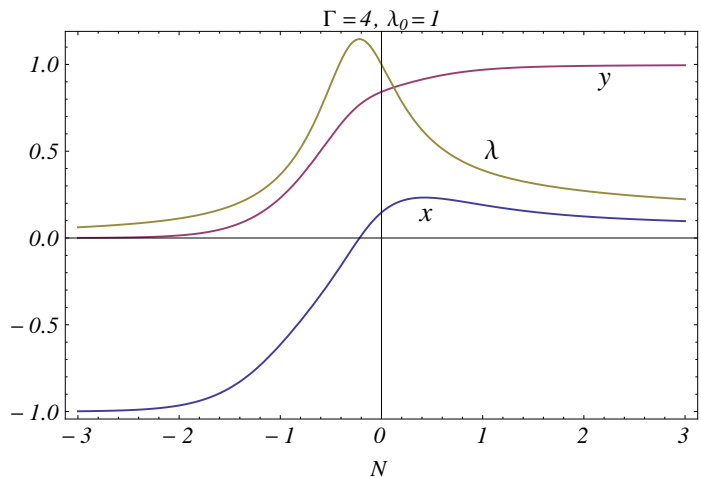}
\caption{The plot of $x,y, \lambda $ vs $N$ for power law type potentials.} 
\end{figure}

\begin{figure}[H]
\centering
\includegraphics[scale=0.5]{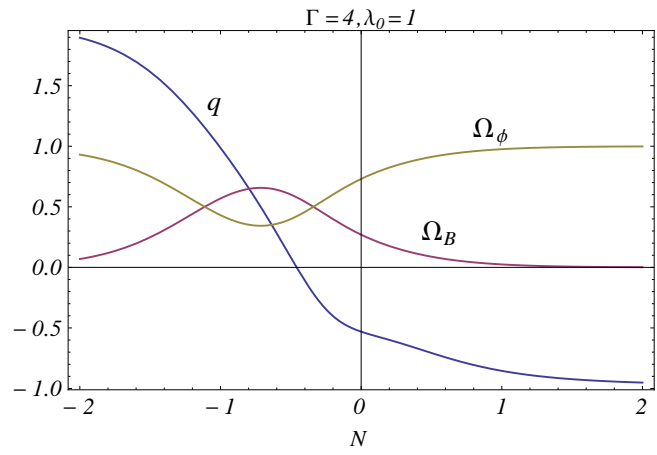}
\caption{The plot of $q,\Omega_{B},\Omega_{\phi}$ vs $N$ for power law type potentials.}
\end{figure}

\begin{figure}[H]
\centering
\includegraphics[scale=0.5]{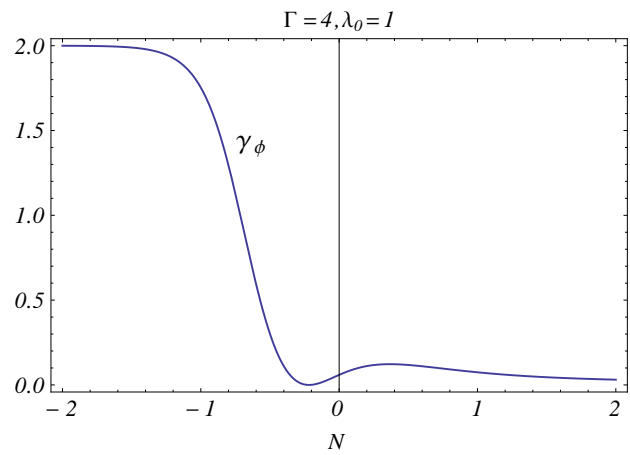}
\caption{The plot of $\gamma_{\phi}$ vs $N$ for power law type potentials.}
\end{figure}

\begin{figure}[H]
\centering
\includegraphics[scale=0.4]{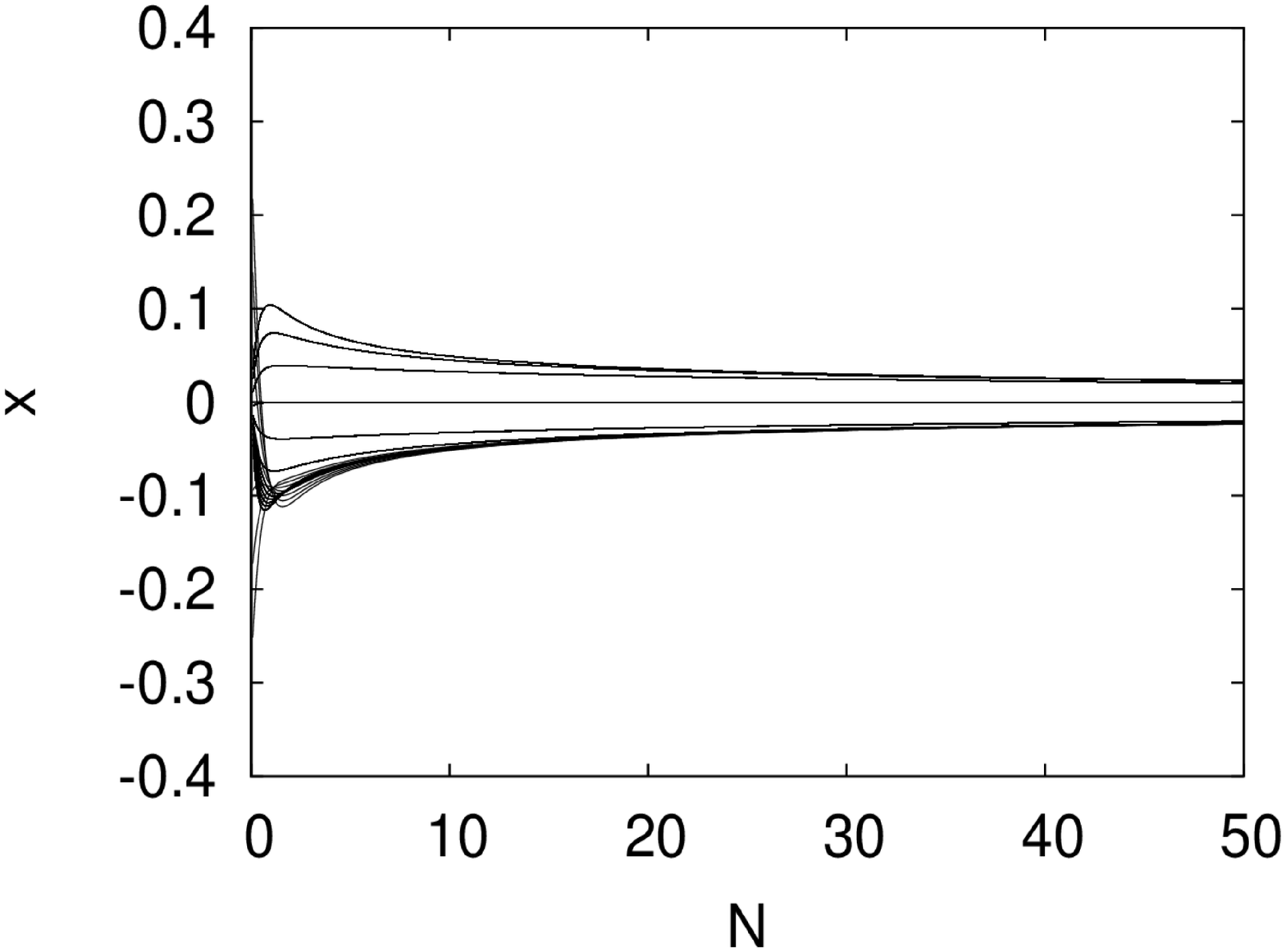}
\caption{The plot of $x$ vs $N$ for perturbation near $(0,1,0)$.} 
\end{figure}

\begin{figure}[H]
\centering
\includegraphics[scale=0.8]{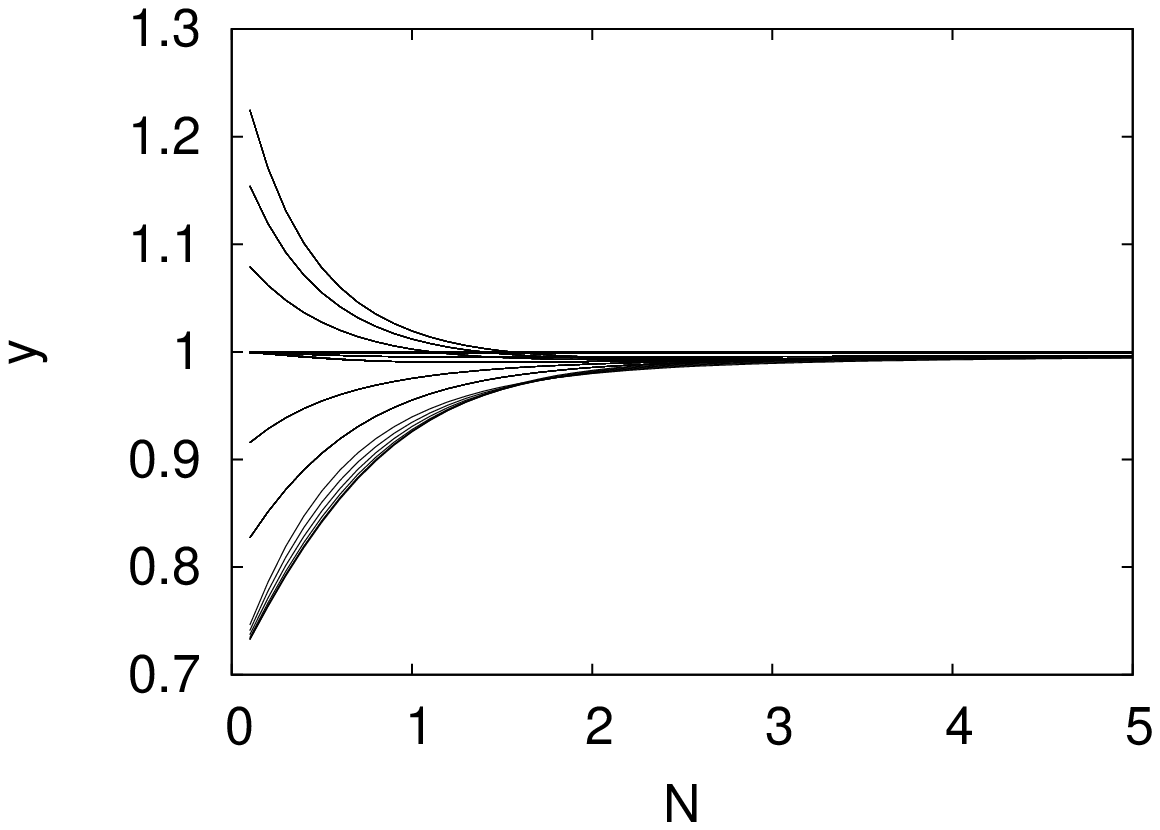}
\caption{The plot of $y$ vs $N$ for perturbation near $(0,1,0)$.} 
\end{figure}

\begin{figure}[H]
\centering
\includegraphics[scale=0.4]{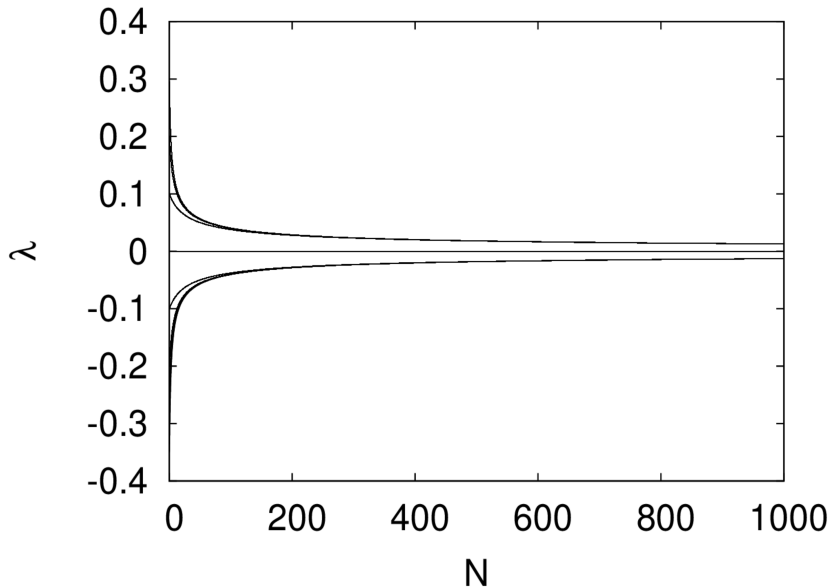}
\caption{The plot of $\lambda$ vs $N$ for perturbation near $(0,1,0)$.} 
\end{figure}

\section{Discussion:}
The present work deals with a dynamical systems study of a quintessence scalar field that drives the recent accelerated expansion of the universe. Two examples of the scalar potential are actually worked out, namely an exponential and a power law type potential. The boundary values are chosen so that they are consistent with the present observations. In both cases, it is possible to find a scenario where the universe in fact starts from an unstable fixed point (hence apt to move away from that)  and evolve to a future attractor through the present phase of accelerated expansion. Amongst the several possible fixed points, only those are chosen which are past and present asymptotes for the set of solutions.\\

The present universe is dominated by the quintessence field, ${\Omega}_{\phi}=0.73$ whereas ${\Omega}_{B}=0.27$. This, in fact, is put into the system as a boundary value. It is intriguing to note that evolution suggests that ${\Omega}_{\phi}$ actually dominates over its dark matter counterpart in the past as well except for a brief interval in the recent past. This feature is there for both the examples, i.e., an exponential or a power law potential. The reason that the universe decelerated in the past is the fact that the equation state parameter ${\gamma}_{\phi}$ had been more than unity so that the quintessence field could not generate the all important negative pressure.\\

Both the examples lead to a situation where the universe in future will be governed completely by the quintessence field (${\Omega}_{\phi}=1$ and ${\Omega}_{B}=0$) and will be steadily accelerating. Although we have worked out for two examples, it appears from equation(\ref{eq.lambda}), the method can, in principle extended for any potential for which $\Gamma$ can be expressed as a function of $\lambda$.
\\

It is important to note that a small change in the initial conditions does not inflict any major or qualitative change in the system. For instance, the system does not show any chaotic behaviour. This has been carefully checked. \\

{\bf Acknowledgments} One of us (N.R.) wishes to thank CSIR (India) for financial support. We thank Prof. Jayanta Bhattacharya for a patient hearing and some very useful suggestions.


\begin{thebibliography}{99}

\bibitem{r1a} S. Perlmutter et al, Bull. Am. Astron.Soc., {\bf 29}, 1351 (1997).\\
\bibitem{r1b} S. Perlmutter et al, Astrophys. J., {\bf 517}, 565 (1999).\\
\bibitem{r1c} J. L. Tonry et al, Astrophys. J., {\bf 594}, 1 (2003).\\
\bibitem{r1d} S. Bridle, O. Lahav, J.P. Ostriker and P.J. Steihardt, Science, {\bf 299}. 1532 (2003).\\
\bibitem{r1e} G. Hinshaw et al, Astrophys. J. Suppl., {\bf 148}, 135 (2003).\\
\bibitem{r1f} A. Kogut et al, Astrophys. J. Suppl., {\bf 148}, 161 (2003).\\
\bibitem{r1g} D.N. Spergel et al, Astrophys. J. Suppl., {\bf 148}, 175 (2003).\\
\bibitem{r1h} C.L. Bennet at al, Astrophys. J. Suppl., {\bf 148}, 1 (2003).
\bibitem{r3a} V. Sahni and A. Starobinsky, Int. J. Mod. Phys. D, {\bf 9}, 373 (2000).
\bibitem{r3b} T. Padmanabhan, Phys. Rep., {\bf 380}, 235 (2003).
\bibitem{r3c} E. J. Copeland, M. Sami and S. Tsujikawa, Int. J. Mod. Phys. D, {\bf 15}, 1753 (2006).
\bibitem{r6} E. J. Copeland, A. R. Liddle and D. Wands, Phys. Rev. D {\bf 57}, 4686 (1998).
\bibitem{r7} R. R. Caldwell, R. Dave and P. J. Steinhardt, Phys. Rev. Lett. {\bf 80}, 1582 (1998).
\bibitem{r8} I. Zlatev, L. M. Wang and P. J. Steinhardt, Phys. Rev. Lett. {\bf 82}, 896 (1999).
\bibitem{r9} A. de la Macorra and G. Piccinelli, Phys. Rev. D {\bf 61}, 123503 (2000).
\bibitem{r10} S. C. C. Ng, N. J. Nunes and F. Rosati, Phys. Rev. D {\bf 64}, 083510 (2001).
\bibitem{r11} P. S. Corasaniti and E. J. Copeland Phys. Rev. D {\bf 67}, 063521 (2003).
\bibitem{r12} R. R. Caldwell and E. V. Linder, Phys. Rev. Lett. {\bf 95}, 141301 (2005).
\bibitem{r13} E. V. Linder, Phys. Rev. D {\bf 73}, 063010 (2006).
\bibitem{r4a} J. Martin, Mod.Phys.Lett.A {\bf23}, 1252-1265,2008.
\bibitem{r4b} V. Sahni, arxiv: astro-ph/0403324.
\bibitem{r17} J. Wainwright and G. F. R. Ellis, \textit{Dynamical System in Cosmology} (Cambridge University Press, 2005. )
\bibitem{r18} A.A.Coley, \textit{Dynamical System and Cosmology} (Springer, 2003)
\bibitem{r19} E. Gunzig, V. Faraoni, A. Figeredo and L. Brenig, Class. Quantum Grav. {\bf 17}, 1783 (2000).
\bibitem{r20} J. Carot and M.M. Collinge, Class. Quantum Grav. {\bf 20}, 707 (2003).
\bibitem{r21} L.A. Urena-Lopez, JCAP {\bf 0509}, 013 (2005).
\bibitem{r22} S. Kumar, S. Panda and A.A. Sen, Quantum Grav. {\bf 30}, 155011 (2013).
\bibitem{r23} S. Sen, A.A. Sen and M. Sami, Phys. Lett B. {\bf 686}, 1 (2010).
\bibitem{r31} W. Fang, H. Tu, J. Huang and C. Shu, arxiv:[1402.4005]
\bibitem{r24} N. Roy and N. Banerjee, Gen. Rel. Grav. {\bf 46}, 1651 (2014).

\bibitem{r25} I. Zlatev and P.J. Steinhardt. Phys.Lett.B, {\bf 459}, 570 (1999).
\bibitem{r26} I. Zlatev, L. Wang and P.J. Steinhardt, Phys. Rev. Lett. {\bf 82}, 896 (1999).
\bibitem{r27} P.J. Steinhardt, L. Wang and I. Zlatev, Phys. Rev.D, {\bf 59}, 123504(1999).
\bibitem{r28} L. Wang, R.R. Caldwell, J.P. Ostriker and P.J. Steinhardt, Astrophys. J., {\bf 530}, 17 (2000).
\bibitem{r30} S.H. Strogatz, \textit{Nonlinear Dynamics and Chaos: With Applications to Physics, Biology, Chemistry and Engineering}; Westview Press, Boulder (2001). 
\bibitem{r29} R. Giostri, M.V. dos Santos, I. Waga, R.R.R. Reis, M.O. Calvalo and B.L. Lago, 
              J. Cos. Astrophysics, {\bf 027}, 1203 (2012).



                
\end{thebibliography}
 \end{document}